\documentstyle[12pt]{article}

\def\gsim{\;
\raise0.3ex\hbox{$>$\kern-0.75em\raise-1.1ex\hbox{$\sim$}}\;
}
\def\lsim{\;
\raise0.3ex\hbox{$<$\kern-0.75em\raise-1.1ex\hbox{$\sim$}}\;
}

\begin{document}
\thispagestyle{empty}

\begin{center}
\Large
{\bf  Resonant Spin-Flavor Precession Solution to the Solar Neutrino
Problem and electron antineutrinos from the Sun}
\end{center}

\vskip 0.3cm

\begin{center}
\large
{\bf A.A. Bykov~$^b$}~\footnote{E-mail: bikov@math380b.phys.msu.su},
{\bf V.Yu. Popov~$^b$}~\footnote{E-mail: popov@math380b.phys.msu.su},
{\bf T.I. Rashba~$^a$}~\footnote{E-mail: rashba@izmiran.rssi.ru},\\
 and
{\bf V.B. Semikoz~$^a$}~\footnote{E-mail: semikoz@izmiran.rssi.ru}\\
\end{center}

\begin{center}
{\it $^a$ The Institute of Terrestrial Magnetism,
Ionosphere and Radio Wave Propagation of the Russian Academy of Sciences,
IZMIRAN, Troitsk, Moscow region, 142190, Russia}
\end{center}

\begin{center}
{\it $^b$ Department of Physics, Moscow State University,
119899, Moscow, Russia}
\end{center}

\begin{abstract}
We have found allowed $\Delta m^2$, $\mu B_{\perp}$, $\sin^22\theta$ --
regions for the Resonant Spin-Flavor Precession Solution (RSFP)~\cite
{Akhmedov} to the Solar Neutrino Problem (SNP) fitted with data of all
acting solar neutrino experiments.

The tipical mass difference region $\Delta
m^2\sim5\cdot10^{-9}-2\cdot10^{-8}~{\rm eV^2}$ varying in dependence on the
mixing $0\leq \sin^22\theta\; \raise0.3ex\hbox{$<$\kern-0.75em\raise-1.1ex%
\hbox{$\sim$}}\; 0.05$ is allowed for some descrete magnetic field regions
in a wide range $30~{\rm kG\; \raise0.3ex\hbox{$<$\kern-0.75em\raise-1.1ex%
\hbox{$\sim$}}\; B_{max}\; \raise0.3ex\hbox{$<$\kern-0.75em\raise-1.1ex%
\hbox{$\sim$}}\; 300~kG}$ for the fixed neutrino transition magnetic moment $%
\mu/10^{-11}\mu_B= \mu_{11}=1$ (free unknown parameter) and for different
regular field profiles. Here $B_{max}$ is the maximum value of a regular
large-scale magnetic field $B(r)$ within the convective zone and the upper
value $B_{max}\simeq 300~{\rm kG}$ is chosen from some known MHD constraints~
\cite{Parker}.

Except of the first allowed range with the lowest field strength $30~{\rm %
kG\; \raise0.3ex\hbox{$<$\kern-0.75em\raise-1.1ex\hbox{$\sim$}}\; B_{max}\; %
\raise0.3ex\hbox{$<$\kern-0.75em\raise-1.1ex\hbox{$\sim$}}\; }$ $70~{\rm kG}$
(for the same $\mu_{11}=1$) the values $B_{max}$ within other descrete
allowed regions depend essentially on the profile $B_{\perp}(r)$ of a
regular magnetic field in the convective zone of the Sun.

For non-zero neutrino mixing ($s_2^2=\sin^22\theta\neq 0$) one can find at
99.9\% CL ($\pm 3\sigma$) an upper limit on the mixing, $s_2^2\leq
s_{2max}^2\sim 0.2$, originated by the non-observation of electron
antineutrinos in the capture reaction $\bar{\nu}_ep\to ne^+$ in the
Superkamiokande (SK) detector ($\Phi_{\bar{\nu}}\; \raise0.3ex%
\hbox{$<$\kern-0.75em\raise-1.1ex\hbox{$\sim$}}\; 0.035\Phi_{\nu_B}$, $%
E_{\nu}\; \raise0.3ex\hbox{$>$\kern-0.75em\raise-1.1ex\hbox{$\sim$}}\; 8.3~%
{\rm MeV}$~\cite{Fiorentini}). This limit occurs also sensitive to the
magnetic field parameter $\mu B_{\perp}$. While at 67\% CL ($\pm \sigma$)
there is another more stringent upper limit $s_2^2\leq s_{2max1}^2\sim 0.05$
coming from inconsistency of gallium experiments with others that is also
close to the result $s_2\; \raise0.3ex\hbox{$<$\kern-0.75em\raise-1.1ex%
\hbox{$\sim$}}\; 0.25$~\cite{Akhmedov1}.

A new allowed region ($\Delta m^2\approx 10^{-7}{\rm eV^2}$, $%
\sin^22\theta\approx0.01$) appears in the case of additional assumption that
the Homestake data are 1.3 times less than the experimental event rate.
\end{abstract}


\newpage

\section{Introduction}

The solar neutrino anomaly is strongly established both from five solar
neutrino experiments~\cite{all} and from the theoretical predictions of the
Standard Solar Model (SSM)~\cite{BP98} confirmed by recent helioseismology
observations~\cite{Berezinsky}. One of the possible solutions to the SNP is
RSFP scenario~\cite{Akhmedov} based on the presence of a non-zero neutrino
transition magnetic moment $\mu=\mu_{ij},~i\neq j$~\cite{Valle} is fully
consistent with all currently available solar neutrino data.

In the present work we are considering the RSFP solution for various models
of regular magnetic fields in the Sun varying three parameters: two
fundamental particle physics ones: (i) the neutrino mass difference $\Delta
m^2$, (ii) the neutrino mixing $\sin^22\theta$ and (iii) the magnetic field
parameter $\mu B_{\perp}(r)$ scaled by the factor $\mu$ with the normalized
magnetic moment $\mu/10^{-11}\mu_B = \mu_{11}$.

In our plots we put $\mu_{11} = 1$ obeying most known constraints on an
active-active neutrino transition magnetic moment including the present
laboratory (reactor) bound, $\mu\; \raise0.3ex\hbox{$<$\kern-0.75em%
\raise-1.1ex\hbox{$\sim$}}\; 1.8\times 10^{-10}\mu_B$~\cite{Derbin}. In
contrast to the case of a Dirac neutrino, for a Majorana neutrino there is
only one more stringent astrophysical constraint coming from the cooling of
red giants $\mu\; \raise0.3ex\hbox{$<$\kern-0.75em\raise-1.1ex\hbox{$\sim$}}%
\; 3\times 10^{-12}\mu_B$~\cite{Raffelt}. Nevertheless, even this exception
does not prevent from our analysis of the RSFP scenario for the Sun since
one can easily to choose the neighboring allowed $B_{\perp}$- regions with
higher values of the magnetic field strength ($B_{max}$) if a more stringent
bound on the neutrino magnetic moment is accepted.

\section{Magnetic fields in the Sun}


It is very little known about magnetic field in the Sun. The MHD models~\cite
{Parker} do not exclude the presence of a significant magnetic field (a few
hundred kilogauss) at the bottom of the convective zone of the Sun.
Moreover, modern dynamo theories forbid large scale magnetic fields in the
central part of the Sun with the strength more than 30 Gauss~\cite{Boruta}.
In the most realistic MHD model of regular magnetic fields in the convective
zone one finds the well-known toroidal field in both hemispheres of the Sun~
\cite{Yoishimura} which has visible traces in forms of magnetic field loops
floating up to active bipolar regions seen on the solar photosphere. The
shape and topology of toroidal fields is very complicated and some
simplifications of their profile is used for different magnetic field
scenarios to solve SNP.

In the present consideration two kinds of magnetic field profiles were used.
First one was simple triangle profile of the magnetic field, second -
''smooth'' profile~(see Fig.1). We have varied magnetic field amplitude from
a few kilogauss up to $300$~kG~\cite{Parker}. Thus, we suppose that magnetic
field has a large scale structure and it can be considered as a regular
field fixed during total observation time for all neutrino experiments.

\section{Master equation}

In the case of non-zero vacuum mixing and non-zero transition magnetic
moment, the neutrino propagation in the solar medium with a magnetic field
can be described by the Schr\"{o}dinger-like $4\times4$ evolution equation
for two neutrino flavors $\nu_e$ and $\nu_\mu$ with two helicities
\begin{equation}
i\left(
\begin{array}{l}
\dot{\nu}_{eL} \\
\dot{\tilde{\nu}}_{eR} \\
\dot{\nu}_{\mu L} \\
\dot{\tilde{\nu}}_{\mu R}
\end{array}
\right) = \left(
\begin{array}{cccc}
V_e -c_2\delta & 0 & s_2\delta & \mu B_+(t) \\
0 & - V_e - c_2\delta & - \mu B_-(t) & s_2\delta \\
s_2\delta & - \mu B_+(t) & V_{\mu} + c_2\delta & 0 \\
\mu B_-(t) & s_2\delta & 0 & - V_{\mu} + c_2\delta
\end{array}
\right) \left(
\begin{array}{c}
\nu_{eL} \\
\tilde{\nu}_{eR} \\
\nu_{\mu L} \\
\tilde{\nu}_{\mu R}
\end{array}
\right)~,  \label{master}
\end{equation}
where $c_2 = \cos 2\theta$, $s_2 = \sin 2\theta$, $\delta = \Delta m^2/4E$
are the neutrino mixing parameters; $\mu = \mu_{ij}$, $i\neq j$, is the
neutrino active-active transition magnetic moment; $B_{\pm} = B_x \pm iB_y$,
are the regular magnetic field components which are perpendicular to the
neutrino trajectory in the Sun; $V_e(t) = G_F\sqrt{2}(\rho (t)/m_p)(Y_e -
Y_n/2)$ and $V_{\mu}(t) = G_F\sqrt{2}(\rho (t)/m_p)(- Y_n/2)$ are the
neutrino vector potentials for $\nu_{eL}$ and $\nu_{\mu L}$ in the Sun given
by the abundances of the electron ($Y_e = m_pN_e(t)/\rho (t)$) and neutron ($%
Y_n = m_pN_n(t)/\rho (t)$) components and the SSM density profile $\rho(t)=$
$250~g/cm^{3}$ $\exp(-10.54t)$~\cite{Bahcall}.

Using the experimental data and errors one can find common regions for the
neutrino mixing and for magnetic field parameters that obey solutions for
all experiments. These regions are named "allowed". All presented plots
except of some cases are made at 95\%~C.L.~($2\sigma$).

\section{Results and discussion}

We have found allowed $\Delta m^2$, $\mu B_{\perp}$, $\sin^22\theta$ --
regions in the case of the RSFP solution to the SNP~\cite{Akhmedov}. These
results are fitted with total rates (Figures 2-7) of all acting solar
neutrino experiments such as Homestake, Kamiokande, SAGE, GALLEX,
SuperKamiokande~\cite{all}:
\[
\begin{array}{||c|c||}
\hline\hline
\mbox{Experiment} & \mbox{ratio DATA/BP98} \\ \hline\hline
\mbox{Homestake} & 0.33\pm 0.032 \\ \hline
\mbox{GALLEX+SAGE} & 0.568\pm 0.076 \\ \hline
\mbox{SuperKamiokande} & 0.470\pm 0.008\pm 0.013 \\ \hline\hline
\end{array}
\]
Here BP98 are the theoretical predictions of the SSM~\cite{BP98}.

All allowed regions have squared mass difference $\Delta m^2\approx
5\cdot10^{-9}{\rm eV^2-2\cdot10^{-8}eV^2}$ (Figures 2-5). In this case
neutrinos have resonant spin-flavor conversions in the convective zone. If
neutrinos would have much bigger mass difference (up to the MSW solution $%
\Delta m^2_{MSW}\approx10^{-5}{\rm eV^2}$) resonant points are deep in the
Sun core (less than $0.3$ of the Sun radius), while no significant magnetic
fields are expected there~\cite{Boruta}.

Allowed regions are some discrete areas in a wide range of magnetic field
amplitude values $B_{max}$ changing from $30$~kG to $300$~kG for the fixed
neutrino transition magnetic moment $\mu $ ($\mu _{11}=1$) and for the
regular magnetic field profiles considered here. Allowed regions are
periodical over the magnetic field amplitude $B_{max}$. This result can been
explained in the framework of the simplest model: constant magnetic field in
the convective zone~\cite{BykovPopov}. In this case the probability of $\nu
_{eL}\rightarrow \nu _{\mu R}$ conversion is
\begin{equation}
P=\frac{(2\mu B_{\perp })^{2}}{(V-\Delta \cos 2\theta )^{2}+(2\mu B_{\perp
})^{2}}\sin ^{2}\left( \sqrt{(V-\Delta \cos 2\theta )^{2}+(2\mu B_{\perp
})^{2}}\frac{\Delta r}{2}\right) ,  \label{analytic}
\end{equation}
where $\Delta =\Delta m^{2}/2E$, $V=V_{e}+V_{\mu }$, $\Delta r$ is the
effective width of the magnetic field region in the convective zone. As one
can see in Fig.1 the effective width of the triangle profile is a little
wider than the effective width of ''smooth'' profile. Hence both the
periodicity of allowed regions (Fig.2 and Fig.3) and their dependence on the
magnetic field profile can be explained from Eq.(\ref{analytic}) .

Let us emphasize that first allowed regions (minimal values of magnetic
fields, $B_{max}\approx50$~kG) both for triangle (Fig.2) and "smooth"
(Fig.3) profiles are very similar, because in the case of small magnetic
fields the probabilities depend poorly on a shape of the profile.

Allowed solutions depend very significantly on the neutrino mixing angle
(Fig.4 and Fig.5). There are two trends. First one is the disappearance of
allowed regions with the increase of mixing angle. The upper limit for
mixing angle is $\sin ^{2}2\theta \approx 0.05$. One concludes that RSFP
solution is possible for small mixing angle cases only~\cite{Akhmedov1}.
Second trend is the disappearance of allowed regions in the case of non-zero
mixing for large magnetic field amplitudes although the small magnetic field
case is still alive.

The non-observation of electron antineutrinos in the capture reaction $\bar{%
\nu}_{e}p$ $\rightarrow $ $ne^{+}$ in the SuperKamiokande (SK) detector ($%
\Phi _{\bar{\nu}}\;\raise0.3ex\hbox{$<$\kern-0.75em\raise-1.1ex\hbox{$\sim$}}%
\;0.035\Phi _{\nu _{B}}$, $E_{\nu }\;\raise0.3ex\hbox{$>$\kern-0.75em%
\raise-1.1ex\hbox{$\sim$}}\;$ $8.3~{\rm MeV}$~\cite{Fiorentini}) gives
additional limit on neutrino mixing angles. The antineutrino SK limit~\cite
{Fiorentini} bounds the mixing ($s_{2}^{2}=\sin ^{2}2\theta \leq 0.2$) at
99.9\% C.L. ($\pm 3\sigma $) (Fig.6). This limit occurs also sensitive to
the magnetic field parameter $\mu B_{\perp }$. The bound is more stringent
for larger magnetic fields than for the small ones.

All above mentioned results are in a good agreement with~\cite
{Akhmedov1,GN,Pulido}.

A new allowed region appears if the additional assumption is taken into
account that the Homestake data are 1.3 times less than the experimental
event rate~\cite[(Homestake)]{all}. In this case ($f_{Cl}=1.3$~\cite
{Fiorentini}) the squared mass difference is $\Delta m^2\approx 10^{-7}{\rm %
eV^2}$ and the mixing is $\sin^22\theta\approx0.01$ (Fig.7).

To conclude the RSFP solution to SNP has a good fit for rates observed in
all acting neutrino experiments. This fit is not worse than for other
solutions such as MSW~\cite{MSW} or vacuum oscillation solution~\cite{vacuum}%
. Since a sizable neutrino transition moment is not forbidden and due to the
existence of efficient solar magnetic field, the RSFP solution is still one
of the possible solutions to SNP.

\section*{Acknowledgments}

Authors thank the RFBR grants 97-02-16501, 99-02-26124, T.I.R. and V.B.S.
thank the INTAS grant 96-0659.

\section*{Figure captions}

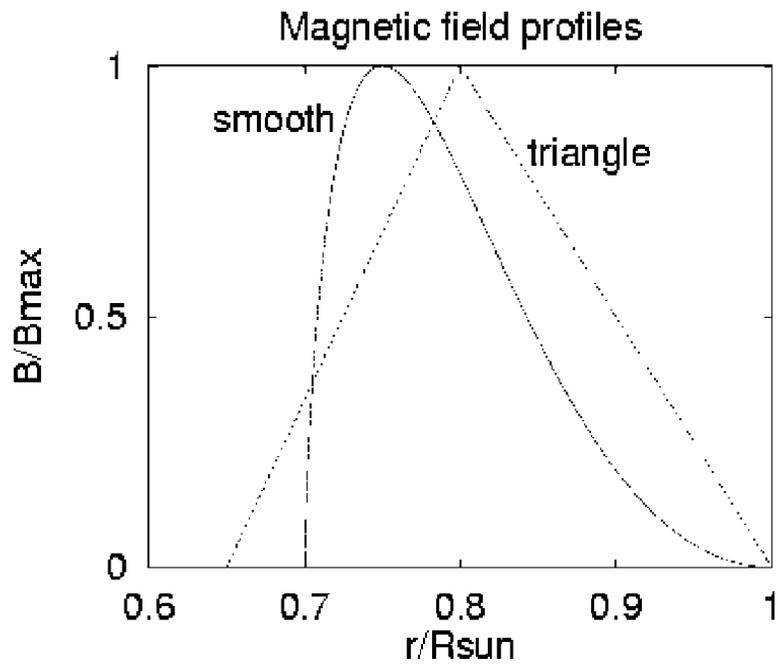
Figure 1. Magnetic field profiles.\newline
Figure 2.~Ratio DATA/BP98 for all acting experiments (Homestake, SAGE,
GALLEX, \mbox{SuperKamiokande}, see Table above) on the plane $\lg\Delta m^2$
(eV$^2$) and $B_{max}/100$~kG for triangle magnetic field profile and zero
mixing angle ($\sin^22\theta=0$) at 95\%~C.L.\newline
Figure 3. Same as Fig.2 for "smooth" profile and $\sin^22\theta=0$.\newline
Figure 4. Same as Fig.2 for triangle profile and $\sin^22\theta=0.05$.%
\newline
Figure 5. Same as Fig.2 for "smooth" profile and $\sin^22\theta=0.05$.%
\newline
Figure 6. Same as Fig.2 for "smooth" profile and $\sin^22\theta=0.2$.
Cross-lined zone is the region excluded by the SK antineutrino bound~\cite
{Fiorentini}.\newline
Figure 7. Same as Fig.2 for triangle profile and $\sin^22\theta=0.01$ and
for the case of the increased Homestake flux $f_{Cl}=1.3$.


\begin{thebibliography}{99}

\bibitem{Akhmedov}  E. Kh. Akhmedov, Phys. Lett. B 213 (1988) 64; C.-S. Lim
and W.J. Marciano, Phys. Rev. D 37 (1988) 1368.

\bibitem{Parker}  E.N. Parker, {\em Cosmological Magnetic Fields}, Oxford
University Press, Oxford, 1979; D.W. Hughes in {\em Advances in Solar System
Magnetohydrodynamics}, ed. by E.R. Priest, Cambridge University Press, 1991;
Ya.A. Zeldovich, A.A. Ruzmaikin, D.D. Sokoloff, {\em Magnetic fields in
astrophysics}, Cordon and Breach, N.Y., 1983

\bibitem{Fiorentini}  G. Fiorentini, M. Moretti and F.L. Villante,
hep-ph/9707097

\bibitem{Akhmedov1}  E.Kh. Akhmedov, ``The neutrino magnetic moment and time
variations of the solar neutrino flux'', Preprint IC/97/49, Invited talk
given at the 4-th International Solar Neutrino Conference, Heidelberg,
Germany, April 8-11, 1997. E.Kh. Akhmedov, A. Lanza and S.T. Petcov, Phys.
Lett. B303 (1993) 85.

\bibitem{all}  K. Lande (Homestake Collaboration) in {\em Neutrino '98},
Proceedings of the XVIII International Conference on Neutrino Physics and
Astrophysics, Takayama, Japan, 4-9 June 1998, edited by Y. Suzuki and Y.
Totsuka, to be published in Nucl. Phys. B (Proc. Suppl.); Y. Fukuda et al.
(Kamiokande Collaboration), Phys. Rev. Lett. 77 (1996) 1683; V. Gavrin (SAGE
Collaboration) in {\em Neutrino '98}; T. Kirsten (GALLEX Collaboration) in
{\em Neutrino '98}; Y. Suzuki (SuperKamiokande Collaboration) in {\em %
Neutrino '98};

\bibitem{BP98}  J.N.Bahcall, S.Basu, M.H.Pinsonnealult, Phys.Lett.B433
(1998)1.

\bibitem{Berezinsky}  V. Berezinsky, astro-ph/9710126, invited lecture at
25th International Cosmic Ray Conference, Durban, 28 July - 8 August, 1997;
V. Castellani, S. Degl'Innocenti, G. Fiorentini, Astron. Astrophys. 271
(1993) 601; W.A. Dziembowski, Bull. Astron. Soc. India 24 (1996) 133; S.
Degl'Innocenti, W.A. Dziembowski, G. Fiorentini, B. Ricci, Astropart. Phys.
7 (1997) 77.

\bibitem{Valle}  J. Schechter, J.W.F. Valle, Phys. Rev. D24 (1981) 1883;
ibid. D25 (1982) 283 (E)

\bibitem{Derbin}  A.V. Derbin, Yad. Fiz. 57 (1994) 236 (Eng. Transl. in
Phys. At. Nucl. 59 (1994) 1171)

\bibitem{Raffelt}  G.G. Raffelt, Phys.Rev.Lett. 64(1990)2856; Phys.Rep.
198(1990)1.

\bibitem{Boruta}  N. Boruta, Astrophys. J., 458 (1996) 832.

\bibitem{Yoishimura}  H. Yoishimura, Astrophys. J., 178 (1972) 863;
Astrophys. J. Suppl. Ser., 52 (1983) 363.

\bibitem{Bahcall}  John N. Bahcall, {\em Neutrino Astrophysics}, Cambridge
University Press, 1988, section 6.3.

\bibitem{BykovPopov}  A.A. Bykov, V.Yu. Popov, A.I. Rez, V.B. Semikoz, D.D.
Sokoloff, hep-ph/9808342.

\bibitem{GN}  M. Guzzo, H. Nunokawa, hep-ph/9810408.

\bibitem{Pulido}  J. Pulido, E.Kh. Akhmedov, hep-ph/9907339.

\bibitem{MSW}  S.P. Mikheev, A.Yu. Smirnov, Sov. J. Nucl. Phys. 42 (1985)
913; Nuovo Cimento C9 (1986) 17; L. Wolfenstein, Phys. Rev. D17 (1978) 2369.

\bibitem{vacuum}  V.N. Gribov, B.M. Pontecorvo, Phys. Lett. B28 (1969) 493.
\end{thebibliography}
\end{document}